\documentclass[fleqn,12pt]{elsarticle}
\biboptions{sort&compress}
\usepackage{graphicx,color}
\usepackage{amsmath,amssymb}
\usepackage{slashbox}
\usepackage{young}
\newcommand{\Slash}[1]{{\ooalign{\hfil/\hfil\crcr$#1$}}}
\newcommand{\tr}{{\rm tr}}

\newcommand{\Nc}{N_{\rm c}}

\newcommand{\lqcd}{\Lambda_{\rm QCD}}

\newcommand{\vp}{\vec{p}}
\newcommand{\vq}{\vec{q}}
\newcommand{\vk}{\vec{k}}

\newcommand{\vr}{\vec{r}}

\newcommand{\la}{\langle}
\newcommand{\ra}{\rangle}

\newcommand{\para}{\parallel}

\newcommand{\calL}{\mathcal{L}}

\newcommand{\calA}{\mathcal{A}}

\newcommand{\rmd}{\mathrm{d}}
\newcommand{\rmi}{\mathrm{i}}
\newcommand{\rme}{\mathrm{e}}

\newcommand{\rmsgn}{ {\rm sgn } }
\newcommand{\tp}{ \tilde{p} }
\newcommand{\tq}{ \tilde{q} }

\newcommand{\tx}{ \tilde{x} }
\newcommand{\ty}{ \tilde{y} }

\begin{document}
%
%
\begin{frontmatter}
\title{The quark mass gap in a magnetic field}
\author{Toru Kojo}
\ead{torujj@physik.uni-bielefeld.de} 
\author{Nan Su}
\ead{nansu@physik.uni-bielefeld.de}
\address{Faculty of Physics, University of Bielefeld, 
D-33615 Bielefeld, Germany}
\address{{\rm (BI-TP 2012/50) } }
\begin{abstract}
A magnetic field and the resulting
Landau degeneracy
enhance the infrared contributions to the quark mass gap.
The gap does not grow arbitrarily, however,
for models of asymptotic free interactions.
For $B \rightarrow \infty$,
the magnetic field decouples from 
the dimensionally reduced self-consistent equations,
so that the gap behaves as 
$\sim \lqcd$ (or less),
instead of $\sim \sqrt{|eB|}$.
On the other hand,
the number of participants to the chiral condensate
keeps increasing as $\sim |e B|$
so that $| \la \bar{\psi} \psi \ra | \sim |e B| \lqcd$.
After the mass gap stops developing,
nothing tempers the growth of screening effects
as $B\rightarrow \infty$.
These features are utilized to interpret
the reduction of critical temperatures for 
the chiral and deconfinement phase transitions
at finite $B$,
recently found on the lattice.
The structures of mesons are analyzed
and light mesons are identified.
Applications for cold, dense quark matter
are also briefly discussed.
\end{abstract}
\end{frontmatter}

\section{Introduction}
In past decades, 
systems in a magnetic field ($B$)
have been useful laboratories 
to test theoretical ideas.
A famous example is a system of cold atoms,
in which a magnetic field controls the strength of the interactions.
In QCD,
similar utilities are also expected for 
the lattice Monte Carlo simulation
at finite $B$ 
\cite{Buividovich:2008wf,D'Elia:2010nq,D'Elia:2011zu,Bali:2011qj}.
In particular, we can study
the nonperturbative gluon dynamics
through polarization effects,
controlling quark dynamics by a magnetic field. 
Such information may help the studies
of cold, dense quark matter \cite{Fukushima:2011jc}.

It seems that lattice studies already
confirmed some of the theoretical ideas.
At $T=0$, a magnetic field enhances
the size of the chiral condensate
due to magnetic catalysis 
\cite{Suganuma:1990nn,Gusynin:1994re}.
A key feature of this phenomenon is
the effective dimensional reduction.
For $B \neq 0$,
the phase space for the low energy particles and anti-particles
is $\sim |eB| \int \rmd p_\para$,
increasing 
the
number of participants to 
the formation of the chiral condensate.
This should be contrast to the $B=0$ case,
where phase space quickly decreases as $\sim \int |p|^2 \rmd |p|$
in the infrared region.
In this case, due to the small number of participants,
the system needs sufficiently strong attractive 
forces to form chiral condensates.

On the other hand,
some surprises have been provided as well 
\cite{Bali:2011qj}.
While $B$ increases the chiral condensate
below the (pseudo-)critical temperatures for
the chiral restoration ($T_\chi$) and deconfinement ($T_{ {\rm D}}$),
those temperatures themselves decrease. 
This might contradict with 
our intuitions,
if we think that a larger chiral condensate
should generate a greater quark mass gap.
Such thinking would suggest that 
(i) a larger quark mass gap should suppress thermal quark
fluctuations, leading to increasing $T_\chi$,
and (ii) a larger quark mass gap
suppresses quark loops, so that
the results should approach
to the pure gauge results,
leading to increasing $T_{ {\rm D}}$.

To resolve this apparent contradiction,
we shall argue that 
the quark mass gap at $T=0$ can
stay around $\sim \lqcd$ (or less) for  large $B$.
Then we can imagine that the gap at $T< T_{\chi, {\rm D} }$
also stays around $\sim \lqcd$,
without strongly suppressing thermal quark fluctuations
and quark loops.
If this is the case, the decreasing of critical temperatures
would not be so unnatural.

In addition,
the aforementioned behavior of the quark mass gap
does not contradict with the growing behavior 
of the chiral condensate, but instead
naturally explains its $B$-dependence at $T=0$.
In fact, the lattice results in \cite{Bali:2011qj} showed the behavior
$\la \bar{\psi} \psi \ra_{T=0}^B \sim |eB| \lqcd$, 
for $|eB| \ge 0.3\, {\rm GeV}^2 \gg \lqcd^2 
\, (\simeq 0.04\, {\rm GeV}^2)$.
Noting that the relation under the Landau quantization,
\begin{equation}
\la \bar{\psi} \psi \ra_{{\rm 4D}}
\sim |eB| \times \la \bar{\psi} \psi \ra_{{\rm 2D}} \,,
\end{equation}
we can see that $\la \bar{\psi} \psi \ra_{{\rm 2D}}$ 
or the quark mass gap 
must be nearly $B$-independent
and $O(\lqcd)$.

In this work we will carry out all the calculations
in the large $\Nc$ limit\footnote{The 
large $\Nc$ limit in a magnetic field
was also studied in Ref. \cite{Fraga:2012ev}
from a different perspective from ours.}.
The use of the large $\Nc$ is motivated by at least three reasons:
(i) At large $\Nc$, gluons are not screened,
so the {\it nonperturbative} forces (i.e. the forces in the infrared)
are stronger than the $\Nc=3$ case.
Such forces can be used to set the {\it upper bound} of the quark mass gap.
(ii) The large $\Nc$ limit has captured
many qualitative aspects of the confined phase at $B=0$.
Therefore it is worth thinking and testing this approximation
in the confined phase at finite $B$,
since its validity and invalidity are not evident
{\it apriori}.
(iii) It is easy to imagine how the $1/\Nc$ corrections
{\it qualitatively} modify the large $\Nc$ results, and 
such corrections just provide welcomed effects
for our scenario (see below).

We will use the large $\Nc$ limit to just claim
that the quark mass gap does not grow much beyond $\lqcd$.
To explain the reduction of the critical temperatures,
in addition we have to argue the $1/\Nc$ corrections.
The quark loops as the $1/\Nc$ corrections 
screen the nonperturbative forces.
As $B$ increases,
the screening effects become larger
because more low energy particles can participate
to the gluon polarization, due to the enhanced Landau degeneracy
$\sim |eB|$ in the 
lowest Landau level (LLL) \cite{Miransky:2002rp}.
If the quark mass gap stops growing
as suggested in our scenario,
there is nothing to suppress the growth of the screening effects 
as $B$ increases\footnote{This suggests that even at $T=0$, 
the nonperturbative gluons will be screened out
at some critical value of $B$ such that the screening mass, 
$m_D \sim g_s |eB|^{1/2} \sim \Nc^{-1/2} |eB|^{1/2} $,
becomes comparable to $\lqcd$
(see also Sec.\ref{conclusion}).}.
Therefore the nonperturbative forces are reduced at large $B$,
and such reduction should lower the critical temperatures
for given $B$.
In addition, hadronic fluctuations as the $1/\Nc$ corrections
also grow as $B$ increases,
helping the chiral symmetry to restore 
\cite{Fukushima:2012xw}\footnote{
For hadronic fluctuations at small $|eB|$, see Ref. \cite{Andersen:2012zc},
where chiral perturbation theory should be at work.}.

We will argue that 
the demanded (nearly) $B$-independent gap of O($\lqcd$)
can be derived, provided that
it is dominantly created by the 
nonperturbative force mediated by the IR gluons.
In particular, both the IR enhancement 
(that is more drastic than the perturbative $1/p^2$ case)
and the UV suppression
of the gluon exchanges are crucial 
for our discussions.
Since we will deal with the LLL which is essentially 
soft physics in the present paper, 
IR enhanced gluon is a key feature 
in this study (see~\cite{Maas:2011se} for a review).
If we include only the perturbative $1/p^2$ force, 
the gap is much smaller than $\lqcd$ 
and depends on $B$ at most logarithmically.
Similar arguments have been used in 
studies of the quark mass function at finite quark 
density \cite{Kojo:2009ha,Kojo:2011cn}.

To illustrate our points,
we first consider the NJL model which does not have
the abovementioned properties.
For $|eB|\rightarrow \infty$,
the gap equation
within the LLL approximation is
\begin{equation}
M_{ {\rm NJL} } (B) 
= G\, \tr \, S(x,x)
\, \longrightarrow \, G\,
 \frac{|eB|}{2\pi} \int \frac{\rmd q_z}{2\pi}
\frac{ M_{ {\rm NJL} } (B) }{ \sqrt{ q_z^2 + M_{ {\rm NJL} }^2 (B) } } 
\, f(q_z, B; \Lambda) \,,
\end{equation}
where $f(q_z, B; \Lambda)$ is some UV regulator function.
The contact interaction couples
all states in the LLL so that
the Landau degeneracy factor $|eB|$
for the LLL appears.
The intrinsic property of the model is that
the chiral condensate has 
the same $B$-dependence
as the mass gap:
\begin{equation}
\la \bar{\psi} \psi \ra_{ {\rm NJL} }^B 
\, \simeq \, -\, \frac{1}{\, G \,} \, M_{ {\rm NJL} } (B) \,.
\label{NJLrelation}
\end{equation}
Depending on the regularization schemes,
$M_{ {\rm NJL} } (B)$ can be 
$\sim |eB|^{1/2}$ 
(proper time regularization \cite{Shovkovy:2012zn}),
or $\sim \Lambda$ 
(four momentum cutoff \cite{Fukushima:2012xw}),
or else.
Each scheme has its own problems.
For schemes predicting the growing behavior of the 
chiral condensate \cite{Skokov:2011ib},
the quark mass gap also develops as $B$ increases. 
Then at finite temperature,
thermal quark contributions are largely reduced
so that
the increasing chiral restoration temperature is naturally expected.
On the other hand, if the mass gap approaches
to constant, the chiral condensate also does,
contradicting with the lattice results.
Therefore, as far as the relation like (\ref{NJLrelation})
is retained, 
it seems that we have to abandon either
the increasing chiral condensate or
the reduction of critical temperatures.

\begin{figure}[tb]
\vspace{0.0cm}
\begin{center}
\scalebox{0.6}[0.6] {
  \includegraphics[scale=.40]{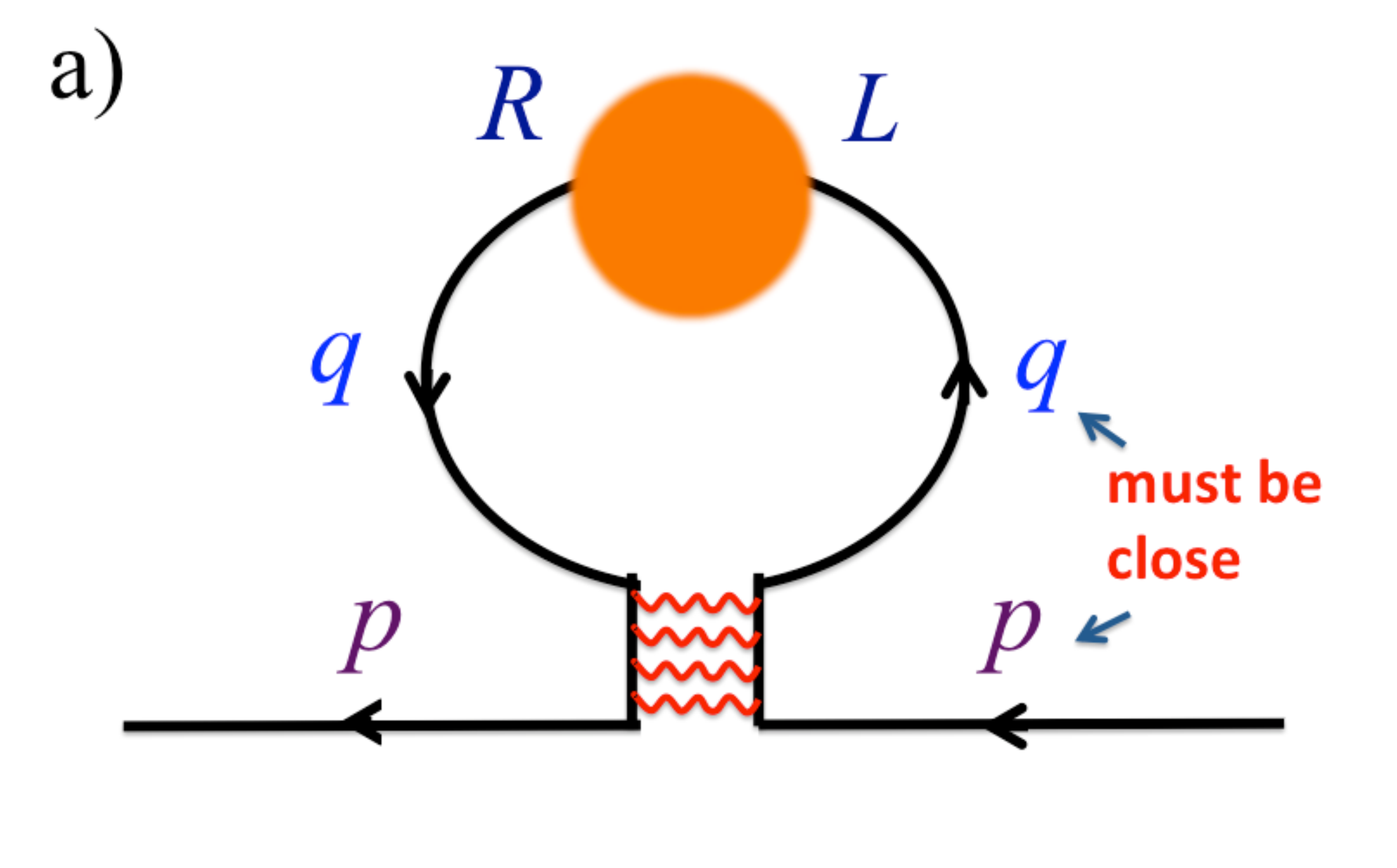} 
} \hspace{0.2cm}
\scalebox{0.6}[0.6] {
  \includegraphics[scale=.40]{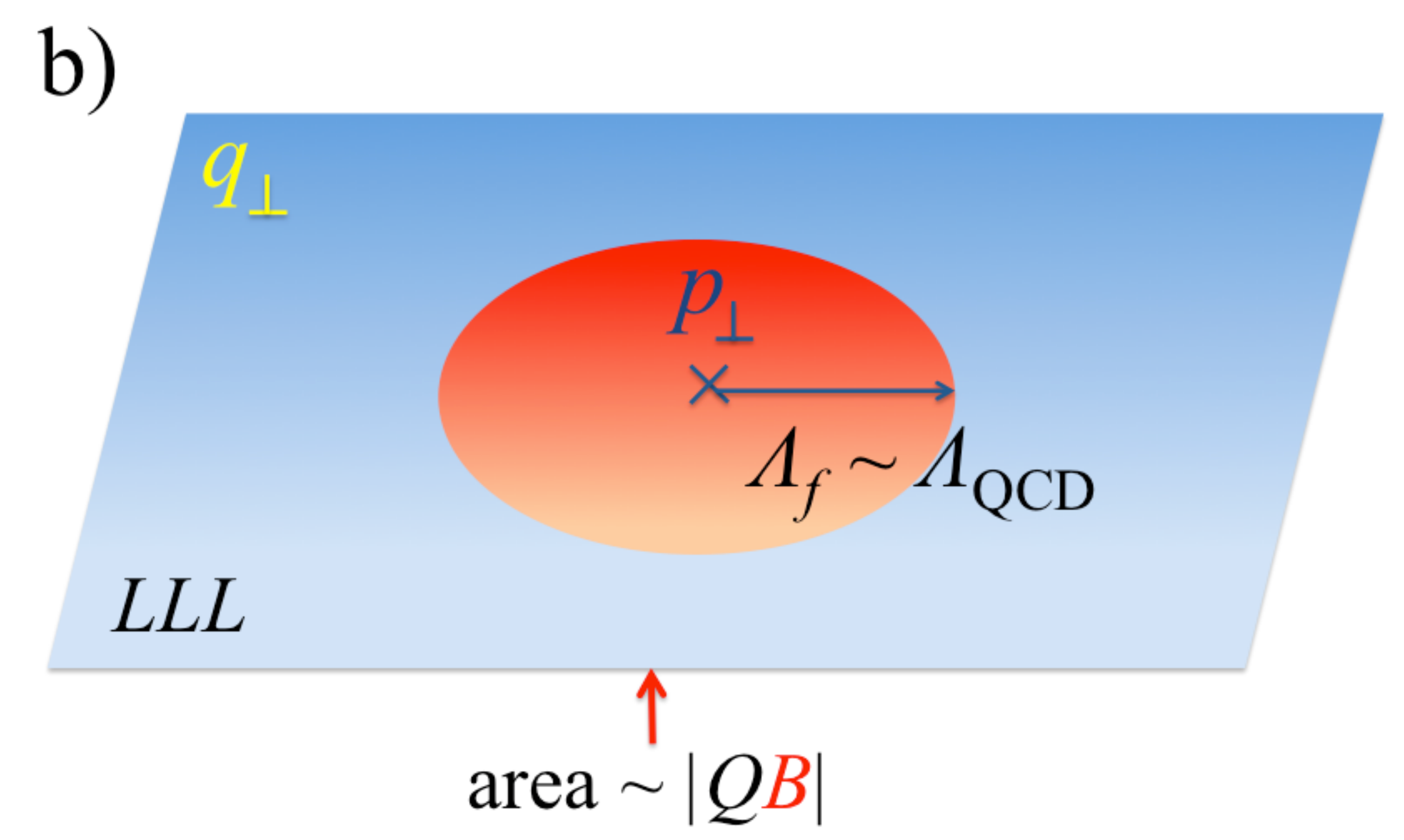} }
\end{center}
\vspace{-0.4cm}
\caption{
(a) The Schwinger-Dyson equation at large $\Nc$
for the model 
in Eq.(\ref{model}).
(b) The distribution of states in the lowest Landau level
for fixed $p_z$.
A state with momentum $p_\perp$ can strongly couple to
states within a domain of 
$|q_\perp - p_\perp| \le \Lambda_f \sim \lqcd$.
}
\label{fig:SDeq}
\end{figure}

This dilemma can be bypassed
if we use the
gluon exchange type interactions with
the IR enhancement and UV suppression.
To emphasize the point,
we consider a simple model for the gluon exchange
with these features
(for the moment we ignore spinor structures),
\begin{equation}
D(q) = G \, \theta( \Lambda_f^2 - \vq^{\, 2} ) \,,
~~~~~~(\Lambda_f \sim \lqcd)
\label{model}
\end{equation}
which was proposed in \cite{Kojo:2011cn}.
In this model, the quark mass function 
appears to be momentum dependent.
For $B \rightarrow \infty$,
the Schwinger-Dyson equation at large $\Nc$ 
(Fig.\ref{fig:SDeq}a) is
\begin{equation}
M (p; B) 
\simeq G \int \frac{ \rmd^4 q}{ (2\pi)^4 } \, 
\tr \, S_{ {\rm 2D}}^{LLL} (q_z) \,  
\theta( \Lambda_f^2 - |\vp -\vq|^2) \,,
\end{equation}
where $B$ is used to reduce the 
quark propagator to the (1+1)-dimensional one
and to separate higher Landau levels
from the LLL.
In contrast to the previous case,
the factor $|eB|$ does not appear in front of the integral.
This is because 
the interaction (\ref{model}) does not couple all the states in the LLL,
but couples the states having similar 
momenta (Fig.\ref{fig:SDeq}b).
This feature makes the gap $B$-independent.
In fact, carrying out
the integral over the transverse momenta, we get
\begin{equation}
M (p; B) 
\, \simeq \,
\frac{G}{2\pi} \int \frac{\rmd q_z}{2\pi}
\frac{ M (q;B) }
{ \sqrt{ q_z^2 + M^2 (q;B)  } } \,
\theta(\Lambda_f^2 - |p_z - q_z|^2) F(p_z-q_z)\,,
\label{damping}
\end{equation}
where $F(k_z) = \sqrt{ \Lambda_f^2 - |k_z|^2 }$.
Note that the equation does not have any explicit $B$ dependence,
so the mass gap is solely determined by the scale $\Lambda_f$, 
i.e. $M (p; B) = M_{\Lambda_f}(p)$.

Another important feature is the damping behavior of
the mass function at large $p$.
To see this, in (\ref{damping}) 
we take $p_z \rightarrow \infty$,
for which $q_z$ must go to $\infty$.
Then the integrand in the RHS goes to zero.
The phase space integral cannot compensate
the damping behavior of the integrand,
because the integral is limited
within the finite domain $\Lambda_f$ around $p_z$.
So $M(p_z) \rightarrow 0$ as $p_z \rightarrow \infty$.
Thus there must be some damping region of $M(p_z)$
which we will denote $\Lambda_\chi$ as 
a function of $\Lambda_f$.
The emergence of the damping scale $\Lambda_\chi$
makes the chiral condensate UV finite\footnote{In vacuum,
$\Lambda_\chi$ effectively plays the role
of the UV cutoff in the NJL model.
}:
\begin{equation}
\la \bar{\psi} \psi \ra
\sim \, - \,|eB| \int^{\Lambda_\chi}_{-\Lambda_\chi}
 \frac{ \rmd q_z}{ 2\pi} \, 
\frac{ M (q_z) }{ \sqrt{ q_z^2 + M^2 (q_z) } }
\, \sim \, -\, |eB| \, F(\Lambda_f)   \,,
\label{confcondensate}
\end{equation}
where $F(\Lambda_f)$ is a function of $\Lambda_f$.
The chiral condensate depends on $B$ {\it linearly},
as advertised.

Note that all of the discussions presented so far
did not ask whether the interaction is confining or not.
All we needed was the IR enhancement.
Nevertheless, it is interesting to
further investigate a model
which not only contains properties of the IR enhancement,
but also
captures certain aspects of the confinement.
For this purpose, we will use the model of Gribov-Zwanziger
\cite{Gribov:1977wm}
which contains the linear rising potential between colored charges.
Confinement is expressed as the absence of the quark continuum 
in the meson spectra.
We will analyze the self-consistent equations for the quark self-energy
and meson states at large $\Nc$ and large $B$,
by reducing them into those of the 't Hooft model.
Then we will rederive the aforementioned conclusions.
%
\section{Dimensional reduction}
\label{DR}
We consider the Euclidean action
(convention:
$g_{\mu \nu} = \delta_{\mu \nu}$, $\gamma_\mu = \gamma_\mu^\dag$):
\begin{equation}
S_E = \int \rmd^4
 x ~ 
\bar{\psi} \left[ \Slash{\partial} + \rmi Q\Slash{\calA} + m \right] \psi 
+ S_{int}
\,,
\end{equation}
where $\calA$ is a $U(1)_{em}$ gauge field,
$Q$ is a flavor matrix for electric charges,
and $m$ is the current quark mass matrix.
The color gauge interaction is treated as $S_{int}$.

We apply an external, uniform magnetic field in 
spatial $3$-direction\footnote{We try to explain only
the lattice results for uniform $B$ in the quenched QED.
If we had some charged condensates
(see recent discussions,
\cite{Chernodub:2011mc,Hidaka:2012mz,Chernodub:2012zx}),
the dynamical QED would generate vortices
in electric superconductor.
},
which can be given by a vector potential
$(\calA_1, \cdots, \calA_4)  = (0, Bx_1, 0, 0)$.
For later convenience, 
it is useful to introduce the projection matrices.
We decompose fermion fields,
\begin{equation}
P^Q_\pm = \frac{1 \pm \rmsgn(QB) \sigma_3}{2} \,,
~~~ \sigma_3 = \rmi \gamma_1 \gamma_2 \,,
~~~ \psi_{\pm} = P^Q_\pm \psi \,,
~~~ \sigma_3 \psi_\pm = \pm \, \rmsgn(QB) \, \psi_\pm \,.
\end{equation}
We expand the fermion fields 
in bases which diagonalize the unperturbed Lagrangian,
\begin{equation}
\psi_\pm (x)
= \sum_{l=0} \int \frac{\rmd^3 \tilde{p} }{(2\pi)^3} \,
\psi_{\pm} ( l,\tilde{p} ) 
\, \rme^{-\rmi \tilde{p} \tx } \, H_l (x_1;p_2) \,,
\end{equation}
where $\tilde{p} \equiv (0, p_2, p_3,p_4)$.
$H_l (x_1;p_2)$ is the harmonic oscillator base
with $m\omega=|QB|$,
whose center is located at $x_1 = p_2/QB$.
If we want to write the Lagrangian as
a sum of the Landau levels,
we should relabel
$\psi_+ (n,\tilde{x} ) \rightarrow \psi_{n+} (\tilde{x}) \,,
~ \psi_- (n-1,\tilde{x} ) \rightarrow \psi_{n-} (\tilde{x}) \,,
~ \psi_+ (0,\tilde{x}) \rightarrow P^Q_+ \chi (\tilde{x}) \,.
$
Now the index $n$ characterizes the Landau level.
The unperturbed action becomes
\begin{equation}
\int \rmd^4 x \, \calL_{ {\rm unpert.} } (x) 
= \int \rmd^3 \tx \left( \calL_0 (\tx) 
+ \sum_{n=1} \calL_n (\tx) \right) \,,
\end{equation}
after carrying out the integral over $x_1$.
The Lagrangian density is 
($\gamma_L=(\gamma_3,\gamma_4)$ 
and $\gamma_\perp =(\gamma_1,\gamma_2)$)
\begin{align}
\calL_0 
= \bar{\chi} (\tilde{x} ) (\Slash{\partial}_L +m) P^Q_+ \chi (\tilde{x} ) \,,
~~~~~
\calL_n 
= \bar{\psi}_n (\tilde{x} )
\left( \Slash{\partial}_L 
+ \rmi \, \rmsgn(QB) \sqrt{2n|QB| } \gamma_2 + m \right) 
\psi_n (\tilde{x} )
\,.
\end{align}
The propagators for Landau levels 
are\footnote{It is convenient to 
attach the Ritus wavefunctions
to the vertices instead of propagators,
because the different Landau orbits couple
only after interactions are turned on.}
\begin{align}
\big\la \chi(\tx) \bar{\chi} (\ty) \big\ra
&= 
\int \frac{\rmd^3 \tp}{(2\pi)^3} \,
P^Q_+ \frac{\rmi}{\, \Slash{p}_L + \rmi m \,}  
\, \rme^{- \rmi \tp (\tx-\ty) } \,,
\nonumber \\
\big\la \psi_n(\tx) \bar{\psi}_n (\ty) \big\ra
&= 
\int \frac{\rmd^3 \tp}{(2\pi)^3}
\frac{\rmi}{\, \Slash{p}_L - \rmsgn{(QB)} \sqrt{2n|QB| } \gamma_2 + \rmi m \,} 
 \, \rme^{- \rmi \tp (\tx-\ty) } \,.
\end{align}
We also expand the color gauge interactions in the Ritus bases
instead of usual Fourier bases \cite{Ferrer:2012zq},
\begin{equation}
\int \rmd^4 x~ \bar{\psi} \gamma_\mu t_a \psi A_{\mu a} (x)
= \sum_{l,l'=0} 
\int \frac{ \rmd^3 \tilde{p} \, \rmd^3 \tilde{q} }{ (2\pi)^6 }~
\bar{\psi} (l,\tilde{p} ) \, \gamma_\mu t_a\,
\psi (l',\tilde{q} ) \, A^{ll'}_{\mu a} (\tp-\tq; p_2,q_2)\,,
\end{equation}
where the gauge field is convoluted with
the harmonic oscillator bases,
\begin{equation}
A^{ll'}_{\mu a} (\tp-\tq ; p_2,q_2) 
= \int \frac{\rmd k_1}{ 2\pi } \,
A_{\mu  a} ( \tp -\tq , k_1) \int \rmd x_1~ 
H_l (x_1;p_2) H_{l'} (x_1;q_2) \, \rme^{-\rmi k_1 x_1} \,.
\end{equation}
For later convenience,
it is useful to prepare the propagator for 
this convoluted form,
\begin{align}
&\hspace{-0.8cm}
\left\la A^{ll'}_{\mu a}(\tp-\tq ; p_2,q_2) \,
A^{j'j}_{\nu b} (\tq-\tp ; q_2,p_2) \right\ra
\nonumber \\
& \hspace{-0.8cm} = 
\int \frac{\rmd k_1 }{ 2\pi } \,
\int \rmd x_1 \int \rmd y_1~ 
H_l (x_1;p_2) H_{l'} (x_1;q_2) \, D^{ab}_{\mu \nu} (\tp -\tq, k_1) \,
H_{j'} (y_1;q_2) H_{j} (y_1;p_2) 
\, \rme^{-\rmi k_1 (x_1-y_1) } 
\,,
\label{A}
\end{align}
where we used 
$\la A_{\mu a}(p) A_{\nu b}(q) \ra = (2\pi)^4 \delta^4 (p+q) D^{ab}_{\mu \nu}(p)$.
Let us note that
the $H_l(x_1;p_2)$ depends on coordinates in the 
combination, $x_1 - p_2/QB$.
We can simultaneously shift 
$x_1-p_2/QB \rightarrow x_1$,
and $y_1-p_2/QB \rightarrow y_1$,
and get
\begin{equation}
H_l (x_1;p_2) H_{l'} (x_1;q_2) 
H_{j'} (y_1;q_2) H_{j} (y_1;p_2)
\rightarrow
 H_l (x_1) H_{l'} (x_1;q_2-p_2) 
H_{j'} (y_1; q_2-p_2) H_{j} (y_1) \,,
\end{equation}
keeping the other part in Eq.(\ref{A}) invariant.
Thus the integral depends on $p_2$ and $q_2$
only through $p_2-q_2$.
This is natural consequence of the gauge invariance,
$\calA_2 = B x_1 \rightarrow B (x_1 - c)$,
that affects the origin of $p_2$ and $q_2$ but
should not affect the final result.

This complicated expression 
can be drastically simplified 
if (i) the interaction shows an IR enhancement around
$\sim \lqcd$, while it damps quickly in the UV,
and (ii) the magnetic field is sufficiently strong,
$|QB| \gg \lqcd$.
The gluon propagator 
rapidly damps for $\tp-\tq$ and $k_1$ 
much larger than $\lqcd$,
cutting off the domain of integral.
Note also that the function $H(x_1)$ contains the Gaussian function,
$\exp[ - |QB| x_1^2/2 ]$,
cutting off the domain of $x_1$ by
$\sim 1/\sqrt{|QB| }$, as far as
we consider not very high Landau orbits.

Assembling all these properties,
we can make replacements at very large $B$, 
\begin{align}
\rme^{-\rmi k_1(x_1 - y_1)} \,
&\rightarrow \, 1\,,
~~~~~~~ \left( | k_1(x_1-y_1) | \sim \lqcd /\sqrt{ |QB| } \ll 1 \right) 
\nonumber \\
H_l (x_1;p_2-q_2) H_{l'} (x_1) \,
& \rightarrow \, H_l (x_1) H_{l'} (x_1) \,,
~~~~~~~ \left( | p_2-q_2 | \sim \lqcd \ll \sqrt{ |QB| }  \right) \,
\label{replacement}
\end{align}
which simplify the Eq.(\ref{A}) as
\begin{equation}
\left\la A^{ll'}_{\mu a}(\tp-\tq ; p_2,q_2) \,
A^{j'j}_{\nu b}(\tq-\tp ; q_2,p_2) \right\ra \,
\longrightarrow \,
\int \frac{\rmd k_1 }{ 2\pi } \, 
\delta_{ll'} \delta_{j' j}\, D^{ab}_{\mu \nu} (\tp -\tq, k_1) \,.
\label{approximation}
\end{equation}
(Note that indices are for the orbital quanta.
Whether they coincide with the Landau level indices
depends on $\gamma$-matrices to which gluons couple,
because $\gamma_\perp$'s flip spins.)

The final expression itself coincides with
a naive expectation and is not surprising.
Ultimately, as $B\rightarrow \infty$ 
a hopping from one Landau orbit
to others (transverse dynamics)
is completely suppressed.
A nontrivial consequence of the asymptotic free theories,
however, is that
the separation of the Landau levels
is achieved at a relatively small magnetic field,
compared to forces like $1/p^2$.
For the $1/p^2$ force,
the damping of the UV force is much milder
so that the LLL and higher levels do not decouple quickly.
The validity of separation is quantified by examining
how the replacements in Eq.(\ref{replacement}) work.

\section{The self-consistent equations at large $\Nc$
within a confining model}

Now we consider the self-consistent equations,
the Schwinger-Dyson 
and Bethe-Salpeter equations.
Below we set $m=0$.
We use the gluon propagator of the
Gribov-Zwanziger type \cite{Gribov:1977wm},
\begin{equation}
D_{\mu \nu}^{ab} (k) 
= \frac{ \delta_{ab}  }{C_F} \, D_{\mu \nu} (k) \,,
~~~
D_{44} (k) = - \, \delta_{44}  \, \frac{8 \pi \sigma}{ (\vk^2)^2 }\,,
~~~ D_{4j} (k) =  D_{ij} (k) = 0 \,.
\end{equation}
which is motivated by Coloumb gauge studies.
$D_{44}$ gives a linear rising potential 
for the color charges, and
we dropped off terms without the strong IR enhancement.
$\sigma$ is a string tension of $O(\lqcd^2)$, 
$C_F=(\Nc^2-1)/2\Nc$ is Casimir for the adjoint representation.

We will assume that
the self-energies are diagonal for each Landau level,
and are translational invariant.
Then the Schwinger-Dyson equation,
after applying our 
approximations (\ref{approximation}),
becomes
\begin{equation}
\big( \Slash{\Sigma}_L + \Slash{\Sigma}_\perp + \Sigma_m \big)
(n, \tilde{p})
= \int \frac{ \rmd^3 \tilde{q} }{ (2\pi)^3 } \,
\bigg[ \,
\rmi \gamma_4  \, 
S(n, \tilde{q};\Sigma ) \,
\rmi \gamma_4 \, \bigg]
\int \frac{\rmd k_1 }{ 2\pi } \, 
 D_{44} (\tp -\tq, k_1) \,,
\end{equation}
that is diagonal for the Landau orbits.

The computations are particularly simple 
for the LLL.
Below we drop the subscript $0$ for the LLL.
The matrix $\gamma_\perp$ drops off
because of the projection operator $P^Q_+$,
so that RHS of the equation does not have
$\gamma_\perp$, meaning that 
$\Sigma_\perp (\tp)=0$ in LHS.
Then the equation looks like
\begin{equation}
\big( \Slash{\Sigma}_L + \Sigma_m \big)
(\tilde{p})
= \int \frac{ \rmd^3 \tilde{q} }{ (2\pi)^3 } \,
 \,
\rmi \gamma_4  \, 
\frac{~ \rmi \big[ \Slash{q}_L -\Slash{\Sigma}_L(\tq) \big] 
+ \Sigma_m(\tq)  ~}
{ \big[ q_L - \Sigma_L (\tq) \big]^2 + \Sigma^2_m (\tq) } \,
\rmi \gamma_4 \,
\int \frac{\rmd k_1 }{ 2\pi } \, 
 D_{44} (\tp -\tq, k_1) \,.
\end{equation}
Recall that the quark transverse momentum 
is gauge dependent in the sense
that $A_2 = Bx_1 \rightarrow B(x_1 -c)$
affects the origin of the transverse momenta
as $p_2 \rightarrow p_2 + QB c$
and $q_2 \rightarrow q_2 + QB c$.
Doing this shift,
the gluon propagator is unaffected while
the modification appears only through
the self-energy as
$\Sigma(p_L, p_2) \rightarrow \Sigma(p_L, p_2 + QBc)$,
and
$\Sigma(p_L, q_2) \rightarrow \Sigma(p_L, q_2 + QBc)$.
This replacement does not change
the structure of the self-consistent equation at all,
so that we obtain the same solutions for the two cases,
$\Sigma(p_L, p_2) = \Sigma(p_L, p_2 + QBc)$,
which means that $\Sigma(p)$ is $p_2$ independent.

As a consequence, we can factorize the equation:
\begin{equation}
\big( \Slash{\Sigma}_L + \Sigma_m \big)
(p_L)
= \int \frac{ \rmd^2 q_L }{ (2\pi)^2 } \,
 \,
\rmi \gamma_4  \, 
\frac{~ \rmi \big[ \Slash{q}_L -\Slash{\Sigma}_L (q_L) \big] 
+ \Sigma_m (q_L)  ~}
{ \big[ q_L - \Sigma_L (q_L) \big]^2 + \Sigma^2_m (q_L) } \,
\rmi \gamma_4 \,
\int \frac{\rmd k_1 \rmd q_2}{ (2\pi)^2 } \, 
 D_{44} (\tp -\tq, k_1) \,.
\end{equation}
The gluon propagator is smeared
by integrating out the transverse momentum,
\begin{equation}
D_{44}^{ {\rm 2D} } (p_z - q_z)
= \int \frac{\rmd k_1 \rmd q_2}{ (2\pi)^2 } \, 
 D_{44} (\tp -\tq, k_1) 
= \int \frac{ \rmd^2 \vk_\perp }{ 4\pi^2 } 
\frac{ -\, 8\pi \sigma }{ \big( (p_z-q_z)^2 + \vk_\perp^2 \big)^2 } 
= - \, \frac{ 2\sigma}{ \,  (p_z-q_z)^2 \, } 
\,,
\end{equation}
which is a confining propagator in (1+1) dimensions.
The form of the Schwinger-Dyson equation
is exactly same as that of 
't Hooft model \cite{'tHooft:1974hx}
in axial gauge, $A_z=0$,
whose solution is known.
The string tension is related to the
two dimensional gauge coupling as
$\Nc g_{ {\rm 2D} }^2 = 4 \sigma$.

The magnetic field disappears from the equation. 
The only role of the magnetic field is
to make the unperturbed quark propagator (1+1)-dimensional
and to separate the LLL from the other Landau orbits.
The only scale in the equation is $\sigma \sim \lqcd^2$.
Using the dressed (1+1)-dimensional 
quark propagator for Eq.(\ref{confcondensate}),
we can get the chiral condensate at finite $B$.

Since the quark self-energies are not always well-defined,
we will also consider the meson states
which are free from any ambiguities.
One can estimate an effective
or constituent quark mass by examining the meson spectra.
To do this, we study the Bethe-Salpeter equation.

We shall move to Minkowski space,
and treat the homogenous Bethe-Salpeter equation
assuming that the total momentum
of a quark and an anti-quark is sufficiently close to the pole.
The equation for a color singlet channel is
\begin{equation}
\Psi(P;q)_{\alpha \beta}^{ff'}
= - \int \frac{ \rmd^4 k}{ (2\pi)^4 }
\big[ \, 
S(k-P) \, \gamma_0 \, \Psi(P;k) \, \gamma_0 \, S(k+P) \,
\big]_{\alpha \beta}^{ff'} D_{00} ( k - q) \,,
\label{BS}
\end{equation}
where $\Psi(P;q)$ is a meson wavefunction
with total momentum $P$ and relative momentum $q$
for a quark and an anti-quark.
The indices $\alpha, \beta\, (f,f')$ are for spinor (flavor) indices.
We can decompose the wavefunction into different
spinor combinations and flavor structures,
\begin{equation}
\Psi_{\alpha \beta}^{ff'} 
= \left(
\Psi_S + \Psi_5 \gamma_5 + \Psi_L \gamma_L + \cdots
\right)_{\alpha \beta} {\bf 1}_{ff'} + \cdots \,.
\label{mesongamma}
\end{equation}
We must apply further decompositions or take appropriate
linear combinations,
because a magnetic field breaks isospin and rotational 
symmetries explicitly.
Below we restrict ourselves to the meson states with
electric charges and spinor combinations
for which we can close the equation only by the LLL.
One can select out such mesons
using the projection operator 
$P^Q_+ =(1+ \rmi \rmsgn{(QB)} \gamma_1 \gamma_2)/2$.
For the two flavor case, the following
structures satisfy the condition,
\begin{equation}
(u\bar{u}\,, d\bar{d}) \otimes ( 1\,, \gamma_5\,, \gamma_L\,, 
\gamma_L \gamma_5\,,
\sigma_{LL'}\,, \sigma_{\perp \perp'} )\,,
~~~~ (u\bar{d} \,, d\bar{u})
\otimes ( \gamma_\perp \,, 
\gamma_\perp \gamma_5\,,
\sigma_{L\perp} )\,.
\label{list}
\end{equation}
For instance, neutral pions remain light,
while charged ones acquire the masses $\sim \sqrt{|eB|}$.
Other examples are vector mesons.
The longitudinal (transverse) component 
of neutral (charged) vector mesons
can remain light, while others 
not\footnote{More careful considerations are necessary
beyond the large $\Nc$ limit where
the annihilation diagrams may contribute.
In this respect
studies of instantons in a magnetic field are important \cite{Basar:2011by}.}.
This observation seems to be consistent with
recent lattice results \cite{Hidaka:2012mz} 
and model calculations \cite{Simonov:2012if}.

Below we consider the neutral scalar component, $\Psi_S$,
as an illustration,
taking only the LLL into account.
As we saw, the dressed quark propagator
is independent of the transverse momentum.
Then we can conclude that $\Psi_S(P;q)$ is independent of
$P_\perp$, because $P_\perp$-dependence appears only 
through $\Psi_S$ so that
different $P_\perp$ gives the same equation
and thereby the same solution.
We can also conclude that $\Psi_S(P;q)$ is independent of
$q_\perp$, because simultaneous shift of momenta,
$q_\perp \rightarrow q_\perp +c$ 
and $k_\perp \rightarrow k_\perp +c$,
does not affect the equation.

Therefore we can again factorize the equation,
\begin{equation}
\Psi_S(P_L;q_L)
= - \int \frac{ \rmd^2 k_L}{ (2\pi)^2 }
\big[ \, 
S(k_L-P_L) \, \gamma_0 \, \Psi_S (P_L;k_L) \, \gamma_0 \, S(k_L+P_L) \,
\big]
\int \frac{ \rmd^2 k_\perp}{ (2\pi)^2 }
 D_{00} ( k - q) \,,
\label{BS2}
\end{equation}
and obtain the Bethe-Salpeter equation in the 't Hooft model.
There is no $B$-dependence in the equation,
so masses of mesons are characterized by
current quark masses and $\lqcd$.
There are an infinite tower of meson spectra,
due to the string oscillations along $z$-direction.

The total momentum of a quark and an anti-quark
is independent of transverse momentum,
meaning that the meson can move freely in the transverse 
direction without costing 
energy\footnote{The meson dynamics is effectively (1+1)-dimensional,
in the sense that the transverse momenta do not affect much
the energy dispersion.
This means that mesons can easily propagate
in the transverse directions.}.
The inclusion of the higher Landau level
can generate the dependence on the transverse momentum,
which is suppressed by a factor of $\lqcd/|eB|^{1/2}$.
This leads to a spatially anisotropic meson 
Lagrangian \cite{Miransky:2002rp}
for which mesonic fluctuations are strong
due to enhanced phase space in the infrared.
It may significantly affect dynamics at finite temperature
\cite{Fukushima:2012kc},
like a system close to the quasi-long range 
order\footnote{
In the strict quasi-long range order,
$\la \bar{\psi} \psi \ra \sim \rho \la \rme^{\rmi \theta} \ra
\rightarrow 0$ due to the IR divergence of the phase fluctuations. 
But the quark mass gap can remain finite
as far as $\rho \neq 0$ \cite{Witten:1978qu}.} 
 \cite{Berezinsky:1970fr}.

\begin{figure}[tb]
\begin{center}
\scalebox{0.8}[0.8] {
  \includegraphics[scale=.3]{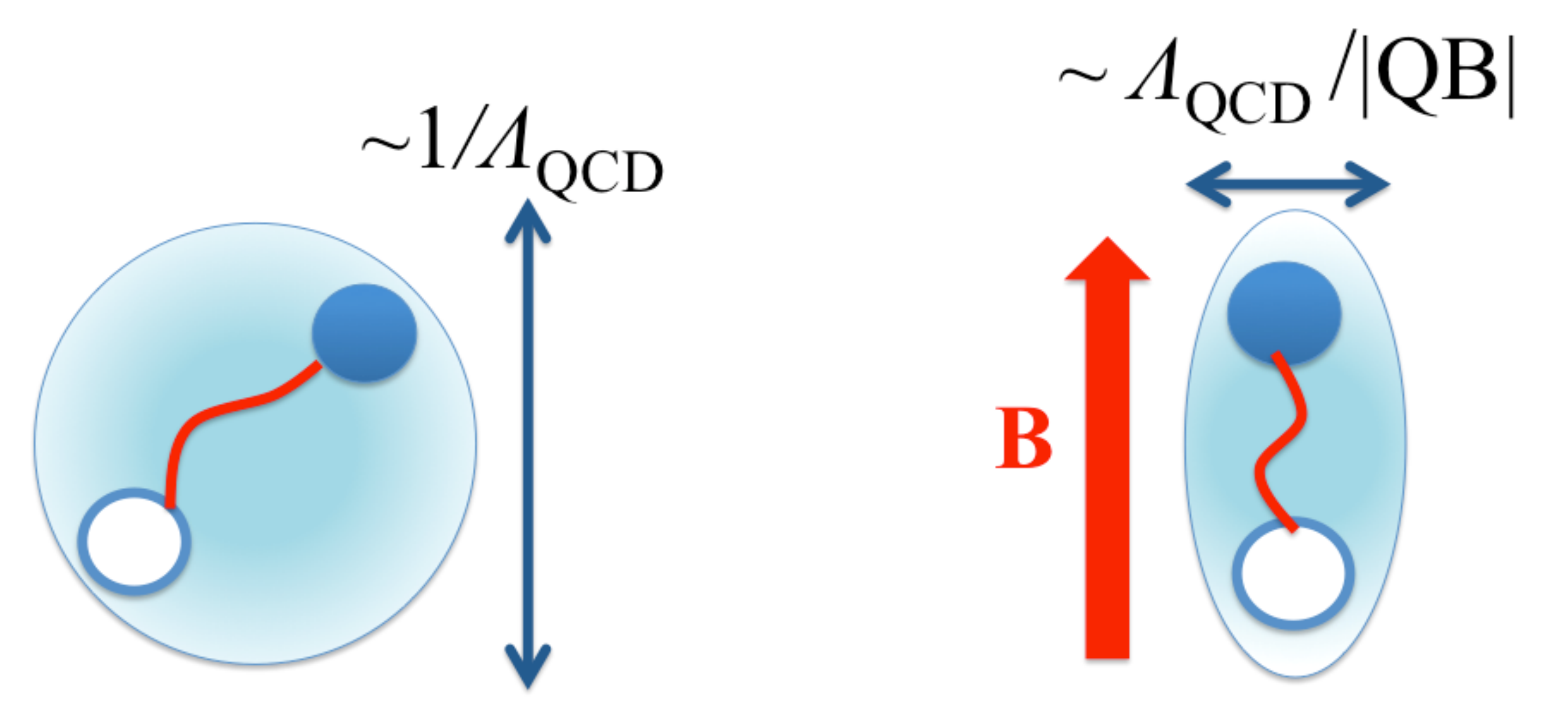} }
\end{center}
\vspace{-0.5cm}
\caption{A structure of mesons listed in Eq.(\ref{list}).
With a magnetic field,
a quark and an anti-quark aligns
along direction of the magnetic field.
The meson can move in the transverse direction
without costing much energy.
}
\label{fig:PT}
\end{figure}

On the other hand, the independence of the relative momentum
means that the internal meson wavefunction behaves as
\begin{equation}
\Psi_S^{ {\rm rel.} } (\vr) 
= \int \rmd q_z \rmd^2 \vq_\perp \, 
\Psi^{ {\rm rel.} }_S (q_z) 
\, \rme^{ \rmi \vq \cdot \vr} 
\sim \delta^2 (\vr_\perp) \, \psi_S(r_z) \,,
\end{equation}
that is, a quark and an anti-quark 
align along the $z$-direction,
and the inter-particle distance is $\sim 1/\lqcd$
(Fig.\ref{fig:PT}).
Couplings with the higher Landau levels
will introduce the width for the transverse wavefunction,
and the mean square radius
should be $\sqrt{ \la \vr_\perp^{\, 2} \ra } \sim \lqcd/|QB|$.

\section{Conclusion}
\label{conclusion}
\begin{figure}[tb]
\begin{center}
\scalebox{0.8}[0.8] {
  \includegraphics[scale=.3]{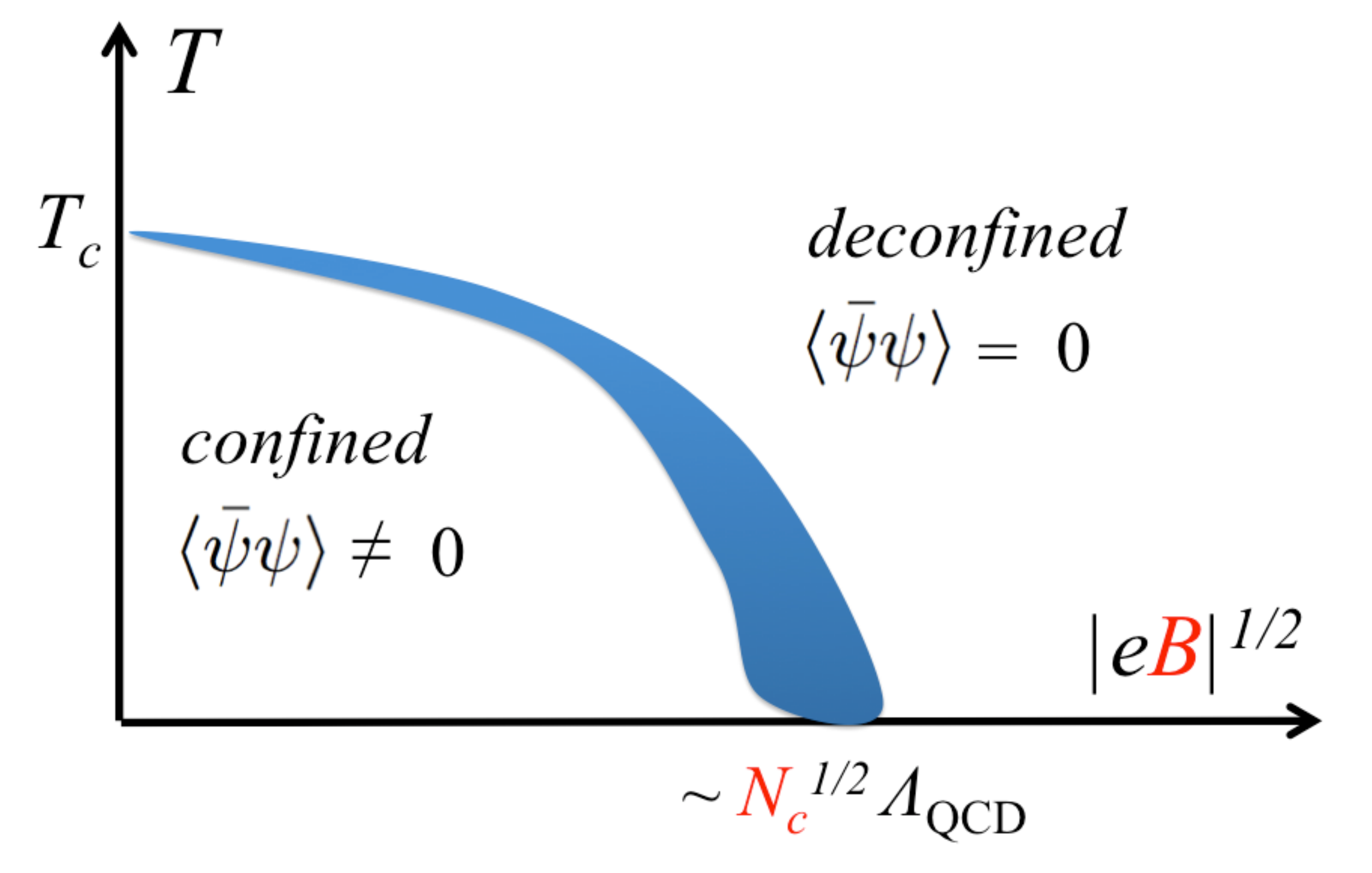} }
\end{center}
\vspace{-0.5cm}
\caption{Our expection for the phase diagram
in $|eB|^{1/2}-T$ plane.
At $T=0$,
a magnetic field would reach its critical strength
around $|eB|^{1/2} \sim \Nc^{1/2} \lqcd$.
}
\label{fig:PT}
\end{figure}
\begin{figure}[tb]
\vspace{0.0cm}
\begin{center}
\scalebox{0.8}[0.8] {
\hspace{0.2cm}
  \includegraphics[scale=.30]{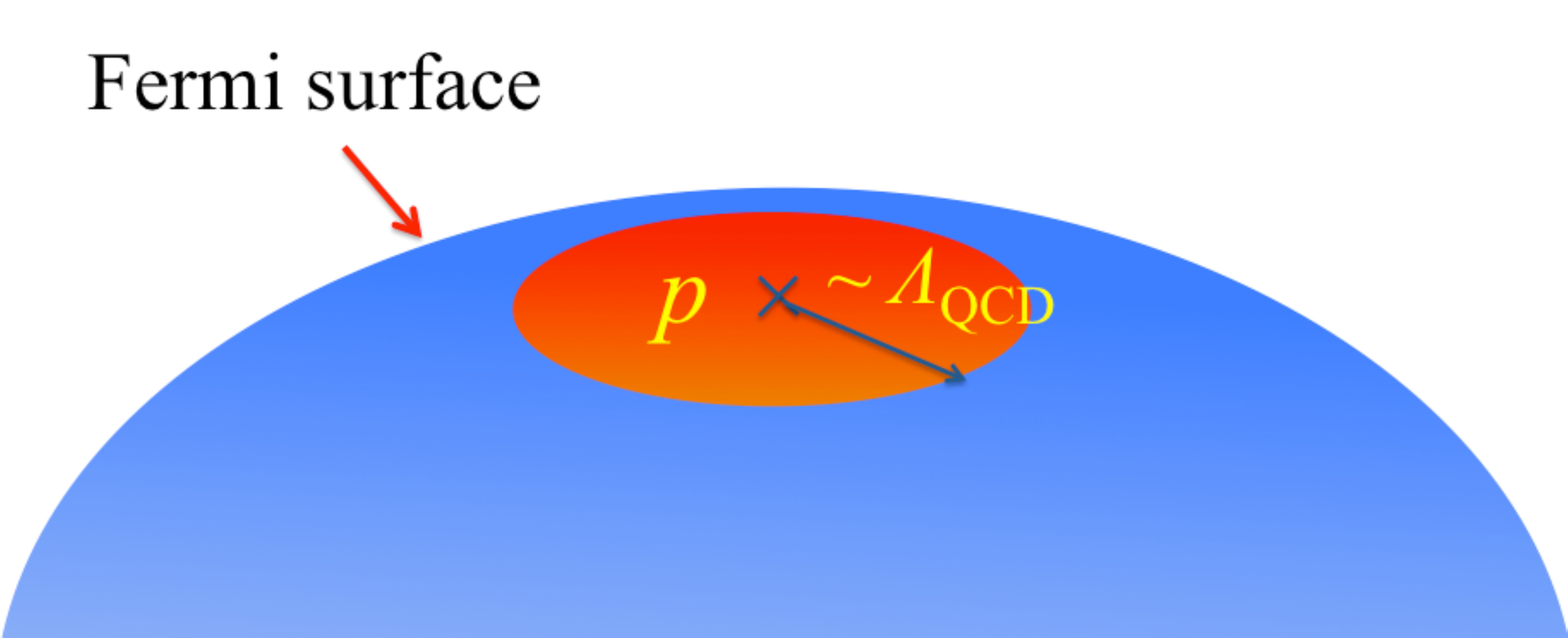} }
\end{center}
\vspace{-0.2cm}
\caption{The momentum space structure 
near the Fermi surface.
If the curvature is small,
the structure of the Schwinger-Dyson equation
for a non-uniform chiral condensate
is very similar to that for the LLL
in a magnetized system
(See also Fig.\ref{fig:SDeq}b).
}
\label{fig:FS}
\vspace{0.2cm}
\end{figure}
We have discussed that the quark mass gap
in the presence of a magnetic field
stays around $\sim \lqcd$ or less,
provided that it is mainly generated
by the nonperturbative part of the gluon exchange.
Accordingly the chiral condensate
grows at most linearly as a function of $B$.
This tendency seems to match with 
the current lattice data.
But to complete our arguments,
more detailed discussions about the UV
tail of the gluon propagator are necessary.
We leave it for future studies.

We have not explicitly taken into account
the screening effects for gluons by taking large $\Nc$.
Including $1/\Nc$ corrections,
the screening effects should grow as $B$ increases,
reducing the nonperturbative forces.
Accordingly the growth rate of the chiral condensate
should become smaller than the large $\Nc$ estimate.
We expect that for an extremely strong magnetic field,
the screening effects will exceed some critical strength
to make the system deconfined
(Fig.\ref{fig:PT}).
For $T=0$, the critical $B$ is roughly estimated 
as $|eB|^{1/2} \sim \Nc^{1/2} \lqcd$, by equating
the numbers of virtual quark and gluon excitations at low energies,
$\Nc |eB| \lqcd \sim \Nc^2 \lqcd^3$.

This situation seems to be quite analogous 
to what would happen in
cold, dense quark matter.
At large quark chemical potential $\mu_q$,
the area of the quark Fermi surface is enhanced.
Accordingly the phase space for quark excitations at low energy
increases as $\sim \Nc \mu_q^2 \lqcd$,
so that the screening effects for gluons become 
stronger\footnote{The phase space enhancement
by $\mu_q$ strongly depends on dimensions.
In particular, in (1+1) dimensions,
the screening does not get stronger.
For QCD in (1+1)-dimensions,
the system is always confined \cite{Schon:2000he}.
}.
The critical chemical potential for deconfinement
has been estimated to be 
$\mu_q \sim \Nc^{1/2} \lqcd$ \cite{McLerran:2007qj}.
Detailed quantitative estimates
crucially depend upon the existence of the quark mass gap
near the Fermi surface.
The gap of $O(\lqcd)$ can emerge
if we take into account the non-uniform chiral condensates.
The corresponding Schwinger-Dyson equation \cite{Kojo:2011cn}
takes very similar form as that in a magnetized system,
Eq.(\ref{damping}).
This correspondence becomes better at larger $\mu_q$,
for which the curvature of the Fermi surface is small enough
(Fig.\ref{fig:FS}).

In summary, the lattice Monte Carlo simulation at finite $B$
is a promising tool to extract out
quantitative information about
the gluon polarizations.
Such information has great relevance to understanding
dense quark matter, or more concretely, physics 
near the Fermi surface.
On the other hand, information about the bulk Fermi sea
should be supplemented
by the lattice studies for the two-color QCD (or isospin QCD)
at large quark (isospin) density without a magnetic field
\cite{Cotter:2012mb}.
These discussions will be expanded elsewhere.

\section*{Acknowledgments}
We thank T. Brauner,
L. McLerran and R. D. Pisarski
for comments and encouragements.
T.K. is supported by
the Sofja Kovalevskaja program and
N.S. by the Postdoctoral Research 
Fellowship of the Alexander von Humboldt Foundation.



\end{document}